\documentclass[english,twoside]{article}
\usepackage{slo,epsfig}
\usepackage{amsmath}
\usepackage{txfonts}
\usepackage{graphicx}
\usepackage{epstopdf}
\usepackage{algorithm}
\usepackage{algorithmic}
\usepackage{comment}
\newtheorem{observ}[theorem]{\protect\observname}
\newtheorem{prop}[theorem]{Proposition}
\newtheorem{lem}[theorem]{Lemma}
\newtheorem{defin}[theorem]{\protect\definname}
\newtheorem{corol}[theorem]{\protect\corolname}
\newtheorem{example}[theorem]{\protect\examplename}

\usepackage{babel}
\providecommand{\definname}{Definition}
\providecommand{\observname}{Observation}
\providecommand{\corolname}{Corollary}
\providecommand{\examplename}{Example}

\def\Z{{\mathbb{Z}}}

\newcommand{\ket}[1]{|#1\rangle}
\newcommand{\bra}[1]{\langle #1|}

\newcommand{\nix}[1]{{}}

\begin{document}
\setlength{\textheight}{8.0truein}    

\runninghead{Efficient approximation of diagonal unitaries over Clifford+T}
            {J. Welch, A. Bocharov, K.M. Svore}

\normalsize\textlineskip
\thispagestyle{empty}
\setcounter{page}{1}

\vspace*{0.88truein}

\alphfootnote

\fpage{1}

\centerline{\bf
Efficient Approximation of Diagonal Unitaries}
\vspace*{0.035truein}
\centerline{\bf over the Clifford+T Basis}
\vspace*{0.37truein}
\centerline{\footnotesize
Jonathan Welch}
\vspace*{0.015truein}
\centerline{\footnotesize\it Department of Chemistry and Chemical Biology, Harvard University}
\baselineskip=10pt
\centerline{\footnotesize\it Cambridge,  Massachusetts, 02138,
USA}
\vspace*{10pt}
\centerline{\footnotesize
Alex Bocharov\footnote{alexeib@microsoft.com}}
\vspace*{0.015truein}
\centerline{\footnotesize\it Quantum Architectures and Computations Group, Microsoft Research}
\baselineskip=10pt
\centerline{\footnotesize\it Redmond, Washington, 98052, USA}
\vspace*{10pt}
\centerline{\footnotesize
Krysta M. Svore\footnote{ksvore@microsoft.com}}
\vspace*{0.015truein}
\centerline{\footnotesize\it Quantum Architectures and Computations Group, Microsoft Research}
\baselineskip=10pt
\centerline{\footnotesize\it Redmond, Washington, 98052, USA}
\vspace*{0.225truein}

\abstracts{
We present an algorithm for the \emph{approximate} decomposition of diagonal operators, focusing specifically on decompositions over the Clifford+$T$ basis, that minimizes the number of phase-rotation gates in the synthesized approximation circuit. The equivalent $T$-count of the synthesized circuit is bounded by $k C_0 \log_2(1/\varepsilon) + E(n,k)$, where $k$ is the number of distinct phases in the diagonal $n$-qubit unitary, $\varepsilon$ is the desired precision, $C_0$ is a quality factor of the implementation method ($1<C_0<4$), and $E(n,k)$ is the total entanglement cost (in $T$ gates). We determine an optimal decision boundary in $(n,k,\varepsilon)$-space where our decomposition algorithm achieves lower entanglement cost than previous state-of-the-art techniques. Our method outperforms state-of-the-art techniques for a practical range of $\varepsilon$ values and diagonal operators and can reduce the number of $T$ gates exponentially in $n$ when $k \ll 2^n$.
}{}{}

\vspace*{10pt}

\vspace*{3pt}
\communicate{to be filled by the Editorial}

\vspace*{1pt}\textlineskip    
\section{Introduction}
Diagonal unitary (DU) operators are used in many quantum algorithms, for example, as simple analytical potential operators for quantum simulation\cite{Zalka:1998eo,Wiesner1996,Kassal:2008uq} and as complex oracles used to divine the answer in quantum searches\cite{GSearch}. The importance of DU operators in a quantum computation tool set has been underscored in \cite{HoggEtAl}.
However, in order to implement a quantum algorithm on a given quantum device, each operator must be decomposed into a sequence of fault-tolerant, device-level instructions.  Therefore, efficient low-level implementation of DU operators is essential to any quantum compiler framework.
In this work we develop methods for the \emph{approximate} decomposition of diagonal operators, focusing specifically on decompositions over the Clifford+$T$ basis.  Since the $T$ gate requires substantial resource overhead as compared to Clifford gates \cite{ZhouLeungChuang,GottChuang} we seek to minimize the number of $T$ gates in our approach.  We analyze the tradeoffs between the $T$-cost of entangling operators and the $T$-cost of decomposing single-qubit rotations.

Methods for the \emph{exact} decomposition of diagonal unitary operators \cite{WelchJ2014,Markov2003} focus on minimization of the number of one- and two-qubit gates,  where the total number of single-qubit axial rotations $R_z$ and two-qubit CNOT operators is upperbounded by $O(2^{n+1}-3)$.
In ~\cite{WelchJ2014}, an $n$-qubit diagonal unitary operator is treated as a discrete function $\{f_k\}_{k=0}^{2^n-1}$.
The discrete function is then amenable to a Fourier-like decomposition over the basis of Walsh functions\footnote{These are a binary periodic functions. See \cite{WelchJ2014} for more details and cf. Figure 1 of that reference for examples of the first eight such functions.}.
A series expansion over this basis provides the following relations
\begin{eqnarray*}
\hat{U}(k)&=&\exp(i f_k)\\
&=&\exp\left(i \sum_{j=0}^{2^n-1} a_j \hat{w}_{jk}\right)\\
&=&e^{i a_0 \hat{w}_{0k}}e^{i a_1 \hat{w}_{1k}}\cdots\\
&=&\prod_{j=0}^{2^n-1} e^{i a_j \hat{w}_{j k}},
\end{eqnarray*}
where $a_j = 1/(2^n)\sum_{k=0}^{2^n-1} f_k \hat{w}_{jk}$.
Each of these basis functions has a one-to-one mapping with a corresponding quantum circuit which implements the basis function as a tensor product of $Z$-rotations on the input register. The splitting of the sum in the exponent into individual exponential terms in the product uses the commutative nature of the $\hat{w}_{jk}$ operators.
The $\hat{w}_{jk}$, in addition to having a functional form, should more generally be thought of as a set of basis circuits.

The primary benefit associated with utilizing the map between the quantum circuits and the corresponding Walsh series expansion is the ability to utilize the tools of Fourier analysis directly on the corresponding DU operator. That is to say, one may allow for a certain error tolerance $\varepsilon$ such that $\left|\hat{U}_\varepsilon(k) - \hat{U}(k)\right|\leq \varepsilon$ for all $k$, and utilize any approximation tools valid for the functional basis as a way of reducing the number of non-zero expansion coefficients that are required for reconstruction of the operator within the specified error tolerance\cite{WelchJ2014}. For DU operators whose discrete functional equivalent $f_k$ has a rapidly converging Walsh series, the corresponding quantum circuit complexity can be reduced significantly as the number of single-qubit rotations (which corresponds exactly with the number of non-zero expansion coefficients in the Walsh series) can often be small and hence efficiently implemented\cite{WelchJ2014}.
These \emph{exact} decompositions of DU operators exhibit the property that all entanglement occurs through the use of elementary CNOT gates, which have negligible fault-tolerant cost.
This results in the entirety of the cost being placed on the fault-tolerant implementation of the set of single-qubit rotations.

In the present work we address two challenges faced in the exact methods. First, there is the challenge that the exact methods are focused on optimizing the complexity of entanglement in exchange for the freedom to require arbitrary rotation angles which are seldom exactly implementable in the Clifford+T basis\cite{ZhouLeungChuang,GottChuang}, and hence single qubit operator approximation methods are required. Secondly, when the phases (diagonal elements) of the operator are sampled from a small set of possible values (e.g., the same phase value appears multiple times on the diagonal), exact methods tend to produce an overly pessimistic number of single-qubit rotations since they rely on treating the diagonal as a discrete 1-D spatial signal and hence depend on the spatial correlations between phase values for small circuit complexity. Here we show that adding one single ancilla qubit we can turn our focus to the domain of phase values, which we call the \emph{phase context} of the operator, and construct networks of entangling operators (referred to as \emph{cascaded entanglers}) whose complexity depends on the number of times a given phase value appears on the diagonal, independent of the T cost required to implement the given phase value as a single qubit rotation.

We present an algorithm to decompose over the phase context of the operator. The rotation angles are approximated and given as the ratio of two distinct phases from the context. This provides the ability to choose one of several possible decompositions so that the required single-qubit rotations can be adjusted to have angles with minimal cost $\varepsilon$-approximations. \footnote{Ancilla-assisted handling of small phase contexts has been proposed as early as \cite{HoggEtAl}. Our approach allows us to reduce the number of required ancillae to exactly $1$.}

The phase-context decomposition requires non-trivial entangling operations which are, in general, multi-controlled Toffoli gates. Though the generalized Toffoli gate can have a potentially significant fault-tolerant cost, the cost is independent of the target accuracy for the single-qubit phase rotation used. Therefore the asymptotic cost of the overall circuit is, in general, dominated by the number of single-qubit phase rotations required. In the case of a \emph{phase-sparse} matrix where the size of the phase context $k$ is much less than the length of the diagonal $2^n$, the small number of single-qubit rotations results in an overall lower fault-tolerant cost for the phase-context approach, despite incurring an additional entanglement cost which turns out to be asymptotically constant. \footnote{An extreme example of ``phase-sparse`` would be a $\mbox{diag}(1,\ldots,1,e^{i \theta})$. Figure \ref{fig:simple:cascading:entangler} shows how to implement it at the cost of one single rotation. The general notion of phase context leads to a generalization of this design.}

It is worthwhile noting that until only recently, it was customary to approximate single-qubit rotations using generic Solovay-Kitaev method (\cite{DN}) that tends to produce approximation circuits with depth scaling like $O(\log^{3.97}(1/\varepsilon))$. The method results in a high number of $T$ gates and using this technique here would favor dramatically the phase-context decomposition that tends to minimize the number of rotations. Recent advances in single-qubit decomposition  (see ~\cite{RoSelinger}, \cite{Selinger}, \cite{BoRoeSvoRUS}, \cite{BoRoeSvoPQF}) provide for effective synthesis of efficient circuits with the $T$-depth in $O(\log(1/\varepsilon))$. The use of these new methods allows us to assume that the $\varepsilon$-dependent part of the cost is of the form $C \, \log_2(1/\varepsilon)$. This results in  more meaningful practical trade-offs between the two diagonal unitary decomposition methods.

The paper is organized as follows.  In Section \ref{sec:phasecontext} phase contexts and cascaded entanglers are defined, and their role in the proposed decomposition is explained.
In Section \ref{sec:ex} we motivate the approach with a simple example. We follow with the general phase-context decomposition algorithm in Section \ref{sec:alg}.
Finally we present numerical results in Section \ref{sec:expts}.

\section{Phase Contexts and Cascaded Entanglers}\label{sec:phasecontext}

Consider a diagonal unitary operator on $n$ qubits with $k \ll 2^n$ distinct phases, such that each distinct phase $\phi_i$ appears in blocks along the diagonal as follows:
\begin{equation}
U=diag(\phi_1,\ldots,\phi_1,\phi_2,\ldots,\phi_2,\ldots,\phi_k,\ldots,\phi_k).
\end{equation}
Throughout, we drop the index on the phase when it is clear from the surrounding text.
We may represent $U$ as a product of a global phase (e.g., $\phi_1$) and a set of one-parameter diagonal operators of the form
\begin{equation}
V(\phi,\ell)=diag(\underbrace{1,\ldots,1}_{2^n-\ell},\underbrace{\phi,\ldots,\phi}_{\ell}).
\label{eq:vphiell}
\end{equation}
The set of $k$ distinct phase values $\Phi = \left\{\phi_i\right\}_{i=1}^k$ is called the \emph{phase context} of the operator. A diagonal operator whose $2^n$ elements are sampled from $\Phi$ requires at most $k-1$ operators of the form $V(\phi,\ell)$ to construct. To see this, suppose the phase $\phi_j$ occurs $\ell_j$ times on the diagonal and suppose $L_m = \sum_{j=1}^{j=m} \ell_j$.
Then, by a direct computation
\begin{equation}
U = \phi_1 \,\, V(\phi_2/\phi_1, L_k - L_1) \,\, V(\phi_3 \, \phi_1/\phi_2, L_k-L_2), \cdot \cdots \cdot V\left(\frac{\prod_{(j=k \text{ mod }2) } \phi_j}{\prod_{(j\neq k\text{ mod }2)} \phi_j}, \ell_k=L_k - L_{k-1}\right)
\end{equation}

We refer to a decomposition of this type as a \emph{phase-context decomposition} (PCD) of the operator $U$. The task is to decompose $U$ into a product of $k-1$ phase rotation gates of the form (\ref{eq:vphiell}).
Suppose $\phi=e^{ i \theta}$, where $\theta \in \mathbb{R}$. We prove that the operator  $V(\phi,\ell)$ can be realized, up to a global phase, using:
\begin{itemize}
\item{One ancillary qubit initialized to $|0\rangle$;}
\item{ A single axial rotation $P(\phi)$ applied to the ancillary qubit; }
\item{ Two identical multi-controlled unitary gates $X^n(V)$ that entangle the primary $n$-qubit register with the ancilla, referred to as  \emph{cascaded entanglers}.}
\end{itemize}

We require several definitions before describing the algorithm.
Let $J=\{|j_1\rangle, \ldots , |j_{\ell}\rangle \}$ be the set of basis vectors rotated by $V=V(\phi=e^{i\theta},\ell)$, where
\begin{equation}
V |j\rangle =
\begin{cases}
\phi  |j\rangle , & |j\rangle \in J \\
|j\rangle , & \mbox{otherwise.}
\end{cases}
\end{equation}
Define $\Omega_\ell(j)$ as the activation function
\begin{equation}
\Omega_J(j) =
\begin{cases}
1 , & |j\rangle \in J \\
0 , & \mbox{otherwise.}
\end{cases}
\end{equation}
Then, with $J=\{\ket{j} : j \geq 2^n-\ell\}$, the operator $V(\phi,\ell)$ can be written as,
\begin{equation}
V = V(\phi,\ell) = \sum_{j=0}^{2^n-1} \left( \phi^{\Omega_{J}(j)} \right) \ket{j}\bra{j}.
\label{eq:Vdef}
\end{equation}

\begin{figure}[htbp]
\centering
\includegraphics[width=3.5in]{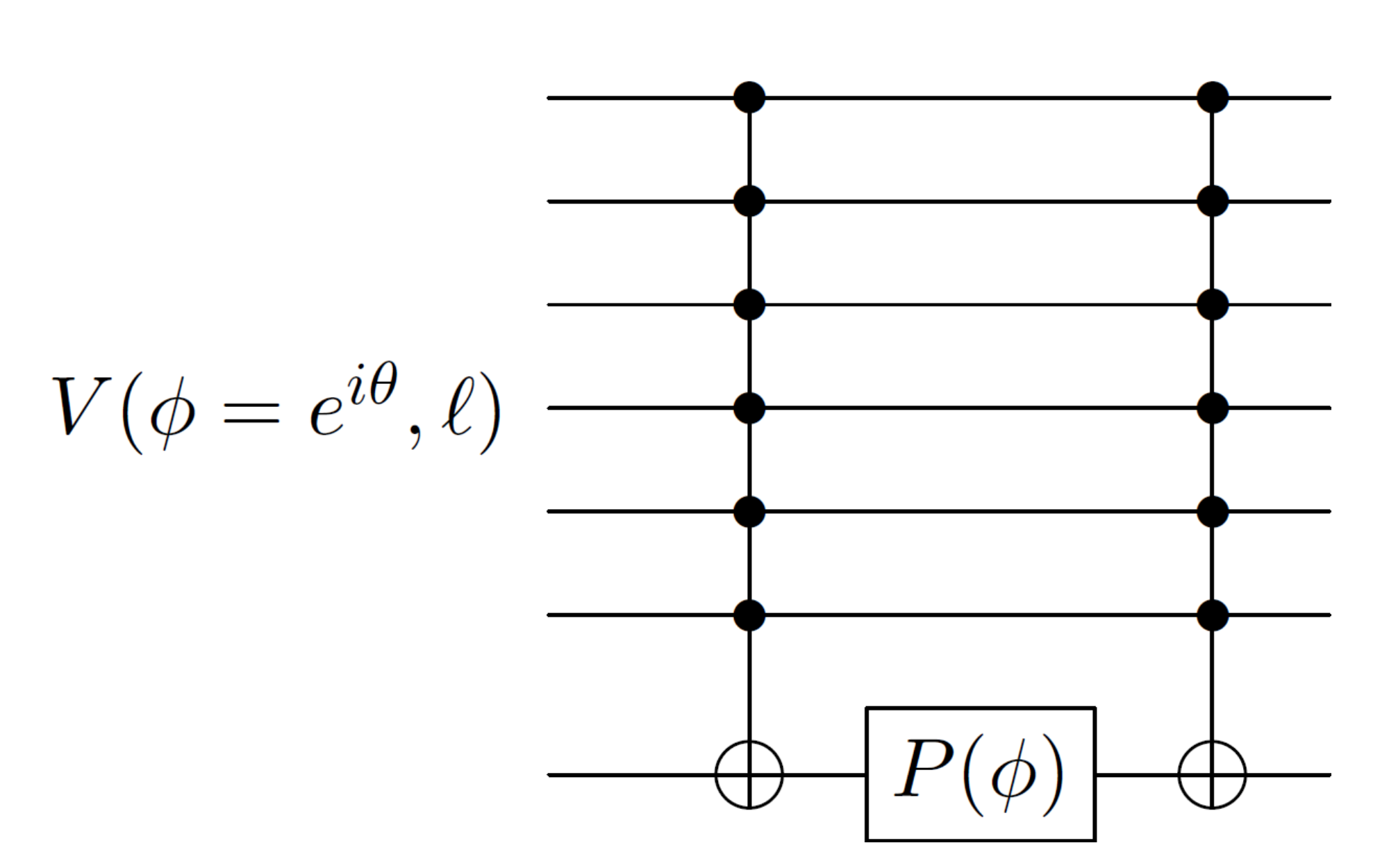}
\fcaption{\label{fig:simple:cascading:entangler}
Circuit to implement $V(\phi = e^{i\theta},\ell=1)$.}
\end{figure}

Each operator $V$ is associated with a so-called \emph{cascaded entangler}, denoted $X^n(V)$, and formally defined on the $(n+1)$-qubit basis as
\begin{equation}
X^n(V)|j\rangle|b\rangle = | j\rangle|b\oplus\,\Omega_J(j)\rangle,
\end{equation}
where $j \in [0,\ldots, 2^n-1] , \, b \in [0,1]$.
Figure \ref{fig:simple:cascading:entangler} shows the the circuit implementation for the case $\ell=1$.

The decomposition of a cascaded entangler $X^n(V)$ results in a circuit whose cost depends only on the number of qubits $n$ and on the structure of the operator $V(\phi,\ell)$, and is independent of the desired precision $\varepsilon$ for approximating the single qubit rotation gate.
Using constructions in Refs.~\cite{GilSel,BarencoEtAl1995}, a cascaded entangler can be represented \emph{exactly} by a Clifford+$T$ circuit where the number of $T$ gates, called the $T$-count, depends only on $n$ and $\ell$.
Moreover, the cost of representing a pair of matching entanglers separated by an ancilla rotation (cf., Figure \ref{fig:simple:cascading:entangler}) is less than twice the cost of one entangler in most cases, since many of the gates in the decompostion of the first entangler cancel with those of the matching entangler.
We denote the minimal $T$-count of a Clifford+$T$ circuit that implements a pair of matching cascaded entanglers as $E[n,\ell]$.

We prove the following statements about cascaded entanglers:
\begin{enumerate}
\item{A cascaded entangler $X^n(V)$ can be represented exactly as a composition of NOTs, CNOTs, and multi-controlled-NOT gates.}
\item{Any multi-controlled-NOT gate on $n$ qubits is represented exactly as a Clifford+$T$ circuit with a $T$-count proportional to the number of controls (using at most one additional ancillary qubit).}
\end{enumerate}


Let $W$ be an arbitrary single-qubit unitary gate:
\[ W= \left( \begin{array}{cc}
w_{00} & w_{01} \\
w_{10} & w_{11} \end{array} \right).\]

Suppose $0 \leq m \leq n , m,n \in \Z$ and consider an arbitrary bit string $b = b_1\ldots b_m$ of length $m$.
Then we define $\Lambda^{n[b]}(W)$ to be an $(n+1)$-qubit gate acting on the $(n+1)$th qubit $t$ as
\begin{equation}
\Lambda^{n[b]}(W) \ket{j_1, \ldots, j_n, t} =
\begin{cases}
w_{t0}\ket{j_1,..., j_{n}, 0 } +  w_{t1}\ket{j_1,..., j_n, 1 } , & \mbox{if} \, \wedge_{k=1}^m (\mbox{XNOR}(b_k, j_k)) = 1, \\
\ket{j_1,..., j_{n}, t} , & \mbox{if} \, \wedge_{k=1}^m (\mbox{XNOR}(b_k, j_k)) = 0,
\end{cases}
\label{eq:xnorop}
\end{equation}
where $\mbox{XNOR}$ is the complement of the exclusive-or operation.
When $W$ is the NOT gate $X$, then
$\Lambda^{n[b]}(X) \ket{j_1, \ldots, j_n, t} = \ket{j_1, \ldots, j_n, t \oplus c_{m[b]}(j)}$, where $c_{m[b]}(j) = \wedge_{k=1}^m (\mbox{XNOR}(b_k, j_k))$.\footnote{$\Lambda^{n[b]}(X)$ is a slight modification of the $N_{[b]}$ notation of \cite{ShendePrasad}. }
As per \cite{GilSel,BarencoEtAl1995} a $\Lambda^{n[b]}(X)$ can be impemented exactly by a Clifford+$T$ circuit with $T$-count in $O(m)$ (assuming an additional ancillary qubit when $n=m+1$).
For simplicity, we drop $[b]$ from the superscript when $m=n$ and $b$ is the bit string of all 1s.

A single-qubit axial rotation $P(\phi)$ can be approximated to precision $\varepsilon$ by a Clifford+$T$ circuit with an expected $T$-count of
$C_0 \log_2(1/\varepsilon) + O(\log(\log(1/\varepsilon)))$, where $C_0$ is a constant depending on the decomposition scheme (cf., \cite{GilSel,BoRoeSvoPQF}).

Thus the $T$-count required to approximate the operator $V(\phi,\ell)$ is bounded by
\begin{equation}
C_0 \log_2(1/\varepsilon) + O(\log(\log(1/\varepsilon))) + E[n,\ell].
\label{eq:asym:tcount}
\end{equation}
The target diagonal operator $U$ is represented as the product
\begin{equation}
U=\prod_{m=1}^{k-1} {V(\phi_m,\ell_m)},
\end{equation}
 and thus can be approximated to precision $\varepsilon$ by concatenating circuits that approximate the respective operators $V(\phi_m,\ell_m)$ to precision $\varepsilon/(k-1)$,
\footnote{Also note that $[V(\phi_i,\ell),V(\phi_j,m)]=0$ and hence the circuits can be implemented in any order.} i.e.,
\begin{equation}
U_\varepsilon=\prod_{m=1}^{k-1}{V_{\varepsilon/(k-1)} (\phi_m,\ell_m)}.
\end{equation}
Therefore the required $T$-count of the overall approximation circuit is bounded by
\begin{equation} \label{eq:multi:phase}
(k-1)  C_0 \log_2(1/\varepsilon) + O(\log(\log(1/\varepsilon))) + E[n,k,\{\ell_1,\ldots,\ell_{k-1}\}],
\end{equation}
where $E[n,k,\{\ell_1,\ldots,\ell_{k-1}\}]$ is an overall  upper bound for the total cost of all cascaded entanglers generated in this decomposition.\footnote{$\sum_{m} \, E[n,\ell_m]$ can be used as a proxy for $E[n,k,\{\ell_1,\ldots,\ell_{k-1}\}]$, although the bound can be tightened in most cases. }

\section{A Motivating Example}
\label{sec:ex}
Let $W \in U(2)$ be an arbitrary single-qubit unitary. Now suppose we intend to implement an $(n+1)$-qubit gate $G=\Lambda^n(W)$.
A traditional approach would be to decompose $G$ into a network of cascaded CNOTs and uncontrolled single-qubit unitaries.
However, unless the resulting single-qubit unitaries can be performed exactly and fault-tolerantly, the major contributor to the asymptotic cost of the decomposition is the cost of approximating the single-qubit unitaries.
When the desired approximation precision $\varepsilon$ is very small, the cost of the single-qubit approximations dominates the total cost of the decomposition of $G$.

Therefore at a small precision level a more cost-efficient approach is to first consider an Euler-angle decomposition of $W$ in order to minimize the number of single-qubit unitaries in the final decomposition.
Let $W=e^{i\delta} \, R_z(\alpha) H  R_z(\beta) H R_z(\gamma),$ where $\alpha, \beta, \gamma , \delta$ are real phase factors and $H$ is the Hadamard gate. Then,
\begin{equation}
\Lambda^n (W) = \Lambda^n(e^{i\delta}) \,\, \Lambda^{n+1}(R_z(\alpha))\,\, \Lambda^n(H) \,\, \Lambda^{n+1}(R_z(\beta)) \,\,\Lambda^n(H) \,\, \Lambda^{n+1}(R_z(\gamma)).
\end{equation}
Recall that $\Lambda^n(H)$ is representable exactly in the Clifford+$T$ basis with a $T$-count of $O(n)$, which is independent of the desired precision \cite{GilSel,AmyEtAl2012}.

The operators $\Lambda^n(e^{i\delta})$, $\Lambda^{n+1}(R_z(\alpha))$, $\Lambda^{n+1}(R_z(\beta))$, $\Lambda^{n+1}(R_z(\gamma))$ are each one-parameter diagonal unitaries.
By allowing one ancillary qubit and using cascaded entanglers, we can approximate each of these uniatries with a circuit whose cost, up to the cost of cascaded entanglers, is dominated by the $T$-count of approximating a single axial rotation.
Assuming the given method for approximating any given axial rotation has an asymptotic cost of the form:
\begin{equation}C_0  \log_2(1/\varepsilon) + O(\log(\log(1/\varepsilon))),\end{equation} we conclude that the cost of implementing $\Lambda^n(W)$ is asymptotically dominated by $4 C_0  \log_2(1/\varepsilon) + O(\log(\log(1/\varepsilon))).$
The associated entanglement cost, which depends only on $n$, becomes increasingly less relevant as $\varepsilon$ tends to $0$.

Of the standard decomposition methods, the technique described in Lemma 7.11 of \cite{BarencoEtAl1995} is suitable to this approximation context. The decomposition is ancilla-assisted and expresses $\Lambda^n (W)$ as a composition of two entanglers and $\Lambda^1(W)$.
Unless the latter is reduced to diagonal unitaries, as proposed above, the decomposition recipes in \cite{BarencoEtAl1995} call for representing $\Lambda^1(W)$ as a circuit consisting of CNOTs and three single-qubit unitaries, where each of the latter may be represented with at most three axial rotations, resulting in a total of nine axial rotations and a corresponding $T$-count upper bounded by $9 C_0  \log_2(1/\varepsilon) + O(\log(\log(1/\varepsilon)))$.
This example demonstrates the approach for the case of (at most) four very special diagonal unitaries, each containing only one non-trivial phase.
In the next section we present the solution for the general case.

\section{Diagonal Operator Decomposition Algorithm}
\label{sec:alg}

We present an algorithm to decompose diagonal unitaries via their phase context, referred to as a \emph{Decomposition with Cascaded Entanglers}, in Algorithm \ref{alg:cascaded:entanglers}.
The algorithm takes as input a diagonal operator $U$, a precision $\varepsilon$, and a method $\mathcal{A}$ for performing single-qubit unitary decompositions, for example using the techniques in Refs.~\cite{Selinger, RoSelinger,BoRoeSvoRUS,BoRoeSvoPQF}.

In Line \ref{l1},  the input diagonal operator $U$ undergoes a phase-context decomposition, where $U$ is represented as the product of a global phase $\phi=e^{i\theta}$ and $k-1$ operators of the form $V(\phi,\ell)$, where $k$ is the number of distinct phase values appearing on the diagonal. In Line \ref{l2} the global phase parameter is saved in \emph{ret}. In Line \ref{l3}, for each of the $k-1$ remaining phases, we store local copies of the operator and the phase gate with which we wish to associate it (Line \ref{l4}), then use the cascaded entangler algorithm to construct the pair of cascaded entanglers associated to the diagonal operator $V$ (Line \ref{l5}).
Line \ref{l6} calls $\mbox{CED2}$ (\emph{cascaded entangler decomposition} for a pair) to obtain an \emph{exact} representation of a pair of matching entanglers over the Clifford+$T$ basis.
The design of the cascaded entangler decomposition (CED) algorithm is described in Section \ref{sec:CPT};
given a gate of the form $X^n(V)$, it decomposes the gate into multi-controlled-NOT gates and Clifford+$T$ circuits
\cite{GilSel,BarencoEtAl1995} for exact representation of each $\Lambda^m(X)$ at a $T$-count in $O(m)$. Line \ref{l7} calls the desired single qubit unitary approximation algorithm to decompose the phase gate $R=P(\phi_j)$. The result is then used as the $\varepsilon$-approximate implementation of $R$ and Line \ref{l8} moves that result in place of $R$ in the circuit for $X^n(V)(I_n\otimes R) X^n(V)$. The latter is an implementation of the individual factor $V(\phi_j,\ell_j)$.
Finally in Line \ref{l9} this result is combined with the global phase calculated earlier and stored in \emph{ret} as the return value for the algorithm.

\begin{algorithm}[H]
\caption{Decomposition with Cascaded Entanglers}
\label{alg:cascaded:entanglers}
\algsetup{indent=2em}
\begin{algorithmic}[1]
\REQUIRE{$n$, $U=diag(\phi_1,\ldots,\phi_{2^n})$, $\varepsilon>0$, $\mathcal{A}$}
\STATE {$factors \gets \phi \prod_{j=1}^{k-1} V(\phi_j,\ell_j)$}\label{l1}
\COMMENT{Phase-context decomposition}
\STATE{$ret \leftarrow \{\phi\}$} \label{l2}
\FOR{$j=1,\ldots,k-1$}\label{l3}
\STATE $V \gets V(\phi_j,\ell_j); R \gets P(\phi_j)$\label{l4}
\STATE $X^n(V) \gets cascaded\_entangler(V)$\label{l5}
\STATE $c_V \gets \mbox{CED2}(X^n(V),I_n\otimes R,X^n(V))$ \label{l6}
\COMMENT{Exact Clifford+$T$ representation}
\STATE $c_R \gets \mathcal{A}(R,\varepsilon)$\label{l7}
\COMMENT{Approximation circuit}
\STATE $c_V \gets replace(R\mapsto c_R \, \mbox{in} \,c_V)$ \label{l8}
\STATE $ret \gets c_V \, ret$ \label{l9}
\ENDFOR
\RETURN{$ret$}
\end{algorithmic}
\end{algorithm}


\subsection{Decomposition of Cascaded Entanglers}
\label{sec:decomp}


We demonstrate that for $V = V(\phi,\ell)$ the cascaded entangler $X^n(V)$ is completely defined by $\ell$ and develop an algorithm for expanding $X^n(V)$ into a network of multi-controlled-NOT gates based on the binary representation of $\ell$.
We then use the network for estimating an upper bound on the cost of $X^n(V)$ over the Clifford+$T$ basis.

Given $\ell \leq 2^n$, let us introduce the $(n+1)$-qubit entanglement operator
\begin{equation}
X^n(\ell) \ket{k}\ket{b}= \left\{
\begin{array}{lc}
\ket{k}\ket{b\oplus 1}&k\geq2^n-\ell,\\
\ket{k}\ket{b}&k<2^n-\ell,
\end{array}
\right.
\label{eq:casc:indicator}
\end{equation} where $\ket{k}$ is an $n$-qubit basis state and $b\in\{0,1\}$.

This operator can be alternatively defined in terms of activation functions as
\begin{equation}
X^n(\ell)=\ket{k}\ket{b\oplus \Omega_{[2^n -\ell,2^n)}(k)}\bra{b}\bra{k}.
\end{equation}

\vspace*{12pt}
\noindent
\begin{observ} \label{observ:key:entangler:fact}
Using a single ancillary qubit in the $(n+1)$th position, $V(\phi,\ell)$ can be implemented as
\begin{equation}X^n(\ell) \, \left(I_n \otimes P(\phi) \right)\, X^n(\ell).\end{equation}
\end{observ}

Indeed, when $X^n(\ell)$ is so defined, then the state $|k\rangle|0\rangle$ picks up the phase factor $\phi$ iff $k \geq 2^n - \ell$, which is exactly how
$V(\phi,\ell) \otimes I$ acts on the state $|k\rangle|0\rangle$.

A suboptimal way of implementing $X^n(\ell)$ would be to factor it into a product $\prod_{j=2^n - \ell}^{2^n - 1} Y_n(j)$, where
$Y_n(j) \, |k\rangle|b\rangle = |k\rangle|b \oplus \delta_{kj}\rangle$, for $ k=0, \ldots, 2^n - 1$ and $b \in \{0,1\}$, is a $\Lambda^n(X)$ gate.
Under this factorization, the cost $X^n(\ell)$ is dominated by $\ell$ times the cost of $\Lambda^n(X)$, which is a uniform worst-case bound.
We show how this bound can be effectively tightened (unless, indeed, it is the definitive worst-case).

\subsection{The Cascaded Entangler Decomposition ($\mbox{CED}$) Algorithm}
\label{sec:CPT}



We propose a simple effective procedure for decomposing the $X^n(\ell)$ operator.
We consider the cost of Pauli and Clifford gates to be negligible and count the number of $T$ gates.
To this end we consider a slightly more general operator $X^n(p,q)$,  where $p  \leq q \leq 2^n$, and the operator is defined in terms of the activation function
$\Omega_{[p,q)}$ as
\begin{equation}
X^n(p,q) \ket{k}\ket{b}= \ket{k}\ket{b\oplus \Omega_{[p,q)}(k) }.
\end{equation}

The gate $X^n(p,q)$, acts to invert the register $\ket{b}, b\in\{0,1\}$, when $p\leq k < q$.
The following identity immediately follows: $X^n(\ell) = X^n(2^n - \ell, 2^n)$.

\vspace*{12pt}
\noindent
\begin{prop}
 $X^n(p,p) = \hat{I}(n)$, where $\hat{I}(n)$ is the n-qubit identity operator for any $p \leq 2^n$.
\end{prop}
\vspace*{12pt}
\noindent
{\bf Proof:}
This follows from Definition (\ref{eq:casc:indicator}) and the fact that $[p,p)$ is empty.
$\square$
\vspace*{12pt}
\noindent
\begin{prop} \label{prop:concat}
For $p < q_1 < q_2$ we have $X^n(p,q_1)X^n(q_1,q_2) = X^n(p,q_2)$
\end{prop}
\vspace*{12pt}
\noindent
{\bf Proof:}
$[p,q_2)$ is the disjoint union of $[p,q_1)$ and $[q_1,q_2)$ and thus
\begin{eqnarray}
\Omega_{[p,q_1)} \oplus \Omega_{[q_1,q_2)}= \Omega_{[p,q_2)}.
\end{eqnarray}

Furthermore,
\begin{eqnarray}
X^n(p,q_1)X^n(q_1,q_2) \ket{k}\ket{b} &=& \ket{k}\ket{b\oplus \Omega_{[p,q_1)}(k) \oplus \Omega_{[q_1,q_2)}(k)}\\
&=& \ket{k}\ket{b\oplus \Omega_{[p,q_2)}(k) }.
\end{eqnarray}
$\square$

\vspace*{12pt}
\noindent
\begin{lem} \label{lem:2:m:block}
Suppose $p=j\,2^m, m \in \Z, j \in \Z, j < 2^{n-m-1}$ and $q = p+2^m = (j+1)\,2^{m}$.
Then $X^n(p,q)$ is an effective \mbox{CNOT}-equivalent of a variably controlled $X$ gate in an $(n+1)$-qubit scheme with $n-m$ levels of control.
\end{lem}

Before proving the lemma we review a convenient set of simple entanglement operators.
Suppose $0 \leq m \leq n,\, m, n \in \Z$.
Consider the operator $\Lambda^{n[j]}(X)$ where $[j]=j_1\ldots j_{n-m}$ is the bit string representation of $j$ possibly padded with zeros.
As defined in (\ref{eq:xnorop}), $\Lambda^{n[j]}(X)$ applied to an $n$-qubit standard basis state $|s\rangle$ executes $X$ on the $(n+1)$-qubit if and only if the first $n-m$ bits of $s$ equal the bit string $[j]$.

\vspace*{12pt}
\noindent
{\bf Proof:}
(Of Lemma \ref{lem:2:m:block})

Consider $j$ which appears in the claim of the lemma. Then $\Lambda^{n[j]}(X) |k\rangle |b\rangle = |k\rangle |b+1\rangle$ if $j \,2^m = p \leq k < j \,2^{m+1} = q$ and
$\Lambda^{n[j]}(X) |k\rangle |b\rangle = |k\rangle |b\rangle$ otherwise.
This is exactly how $X^n(p,q)$ is defined for the particular $p$ and $q$ of the lemma and thus $X^n(p,q)=\Lambda^{n[j]}(X)$.
$\square$

\vspace*{12pt}
\noindent
\begin{corol} \label{corol:X:n:hemming}
Let $h$ be the Hamming weight of $\ell < 2^n$ and let
$\ell=2^{k_1}+\ldots+2^{k_h}$, $k_r \in \Z$, for $r=1,\ldots,h$, and $k_1 < \ldots < k_h$ be standard binary decomposition of $\ell$. Then $X^n(\ell)$ can be effectively factored into a composition of exactly $h$ entanglers of $\Lambda^{n[j]}$ type and the total number of control levels across all these entanglers is exactly $h\,n-(k_1+\ldots+k_h)$.
\end{corol}

\vspace*{12pt}
\noindent
{\bf Proof:}
Consider the following thresholds: $p_r=2^n - \sum_{s=r}^h 2^{k_s}$, $r = 1,\ldots,h$.
Clearly, each $p_r$ is divisible by $2^{k_r}$. More specifically, $p_r = j_r \, 2^{k_r}$ where $j_r = 2^{n-k_r}-\sum_{s=r}^h 2^{k_s-k_r}$.
For convenience, let $p_{h+1} = 2^n$. By design, $p_{h+1} = p_h + 2^{k_h}$.
Also note that $p_1=2^n-\ell$.

Using  Proposition \ref{prop:concat}, we have
\begin{equation} \label{eq:X:n:decomp}
X^n(\ell) =X^n(2^n - \ell, 2^n)=X^n(p_1, p_{h+1})=\prod_{r=1}^h X^n(p_r, p_{r+1}).
\end{equation}

By definition $p_{r+1}=p_r+2^{k_r}=(j_r+1)\,2^{k_r}$, so $X^n(p_r, p_{r+1})$ satisfies the premise of Lemma \ref{lem:2:m:block} (with $m=k_r$).
Therefore each $X^n(p_r, p_{r+1})$ is equivalent to an entangler of $\Lambda^{n[j]}(X)$ type with $n-k_r$ levels of control. There are exactly $h$ entanglers in the product in (\ref{eq:X:n:decomp}). Summing up the control levels completes the proof of the corollary.
$\square$

This corollary provides the most straightforward way of expressing a cascaded entangler in terms of generalized Toffoli gates without the use of additional ancillae. Using known networks of representing generalized Toffolis in terms of three-qubit Toffoli gates \cite{BarencoEtAl1995,ShendePrasad} we observe that the number of three-qubit Toffoli gates in the desired circuit can be made roughly proportional to the total number of levels of control $h\,n-(k_1+\ldots+k_h)$ in the context of Corollary \ref{corol:X:n:hemming}.
To represent $X^n(\ell)$ exactly in terms of the Clifford+T basis we further observe that ultimately the $T$-count of such a representation is going to be roughly proportional to the total number of control levels.
This may be satisfactory in an asymptotic sense, however it turns out to be non-optimal in a practical sense.

For example, the decomposition of $X^5(15)$ as per Corollary \ref{corol:X:n:hemming} yields four $\Lambda^{n[j]}(X)$ type entanglers with a total of $14$ levels of control.
However there is an alternative decomposition with only two entanglers and $6$ levels of control.
Indeed, we may verify $X^5(15)=X^5(2^4,2^5)\, X^5(2^4,17=2^5-15)$.

An empirical optimization of the decomposition of $X^n(\ell)$ can be best described in terms of a `signed bit binary expansion' of $\ell$:
\begin{equation} \label{eq:whatsa:name}
\ell = \sum_{k=0}^K \ell_k \, 2^k, \, \ell_k \in \{-1,0,1\}.
\end{equation}
Such a decomposition is not unique. The following is a recursive description of the specific desired form, denoted $\mbox{bsb}$ for \emph{balanced signed bit}. It is defined as follows:
\vspace*{12pt}
\noindent
\begin{defin}{(Recursive definition of the $\mbox{bsb}$ representation.)} \label{defin:whatsa:name}
Set $\mbox{bsb}(0)=0$ , $\mbox{bsb}(1)=(+1)\,2^0$.

For a given integer $\ell>1$ consider $m=\lfloor \log_2(\ell) \rfloor$.
If $\ell < \frac{4}{3} \, 2^m$, then set $\mbox{bsb}(\ell) = \mbox{bsb}(\ell-2^m)+2^m$; otherwise set $\mbox{bsb}(\ell)=-\mbox{bsb}(2^{m+1}-\ell)+\,2^{m+1}$.
\end{defin}
We note that this definition is designed so that $\mbox{bsb}(\ell)$ tail-recurses with an argument that is smaller than $\ell/2$.
Indeed, by definition $2^m \leq \ell < 2^{m+1}$. If $\ell < \frac{4}{3} \, 2^m$ we use $\ell-2^m < \frac{1}{3} \, 2^m < 2^{m-1} \leq \ell/2$.
If $\ell > \frac{4}{3} \, 2^m$, we use $2^{m+1} - \ell < \frac{2}{3} \, 2^m = (\frac{4}{3} \, 2^m)/2 < \ell/2$. Thus the tail-recursion based on Definition \ref{defin:whatsa:name} is at most $\log_2(\ell)$ deep.

With slight abuse of terminology we call the number of non-zero coefficients in the expansion (\ref{eq:whatsa:name}) the \emph{Hamming weight} of that expansion.
In particular we call the Hamming weight of $\mbox{bsb}(\ell)$ the \emph{balanced Hamming weight} of $\ell$.

Balanced signed bit expansion provides an alternative decomposition of cascaded entanglers that is practically more efficient in a significant majority of cases, as described in the following:
\vspace*{12pt}
\noindent
\begin{lem} \label{lem:X:n:balanced}
Let $h=h(\ell)$ be the balanced Hamming weight of $\ell < 2^n$. Let $\ell=\ell_1 \, 2^{k_1}+\ldots+ \ell_{h} \, 2^{k_{h}}$, be the balanced signed bit decomposition of $\ell$, with $\ell_i \in \{-1,1\}$, for $i=1,...,h-1$, $\ell_h=1$, and $k_r \in \Z$, with $k_r < k_{r+1}$ for $r=1,...,h$.

Then $X^n(\ell)$ can be effectively factored into a composition of exactly $h$ entanglers of $\Lambda^{n[j]}(X)$ type and the total number of control levels across all entanglers is exactly $h\,n-(k_1+\ldots+  k_{h})$.
\end{lem}

\vspace*{12pt}
\noindent
{\bf Proof:}
The proof is almost identical to that of Corollary \ref{corol:X:n:hemming}. Factorization (\ref{eq:X:n:decomp}) also still holds with slightly different thresholds.
Namely, we set $p_r=2^n - \sum_{s=r}^h \ell_s \,2^{k_s}, r = 1,\ldots,h$.
Equation (\ref{eq:X:n:decomp}) works correctly because the inverter gate $X$ on the $(n+1)$th qubit is involutive: $X^2 = I$.
$\square$

The balanced decomposition of $X^n(\ell)$, based on Lemma \ref{lem:X:n:balanced}, leads to more efficient circuits for the majority of values of $\ell$.
Therefore both methods (of Corollary \ref{corol:X:n:hemming} and Lemma \ref{lem:X:n:balanced}) should be used to determine the best circuit decomposition.

We propose an upper bound on the $T$-count for both types of decomposition of $X^n(\ell)$ developed above.
We start by counting the number of three-qubit Toffoli gates required for the decomposition, denoted by $\mathcal{T}$ for convenience.
Note that we can always ensure $m < n$, i.e., $m \leq ((n+1)-2)$, by using an ordinary decomposition (Corollary~\ref{corol:X:n:hemming}) whenever $\ell > 2^{n-1}$.
For simplicity, we focus on the case of $n \geq 4$ (improvements for $n < 4$ may be available on a case-by-case basis).

In the $(n+1)$-qubit register, a constituent entangler of $\Lambda^{n[j]}(X)$ type with $m$ levels of control is Clifford-equivalent to a generalized Toffoli gate with $m$ levels of control.
This is a CNOT gate when $m=1$ and a single three-qubit Toffoli gate when $m=2$.
According to \cite{BarencoEtAl1995} Section 7.1, for $m>2$ the desired generalized Toffoli can be implemented exactly with
\begin{enumerate}
\item{$\mathcal{T} = 4(m-2)$, when $3 \leq m \leq \lceil (n+1)/2 \rceil$;}
 \item{$\mathcal{T} = 8 (n-4)$, when $m=n-1$ and $n \geq 6$.}
 \end{enumerate}

The intermediate values of $m$, $\lceil (n+1)/2 \rceil < m < n-1$, might require separate treatment, however, for the purposes of our cost estimate we disregard some available qubits when $m\geq 5$. Then consider a subregister with $(n'+1)$ qubits where $n' = m+1$ and follow option 2 above.
To summarize, assuming $m$ levels of control, we estimate $\mathcal{T}$ to be $4(m-2)$ when $m \leq \lceil (n+1)/2 \rceil$ and bound it by $8(m-3)$ otherwise.

\begin{figure}[htbp]
{
  \centering
\includegraphics[width=5in]{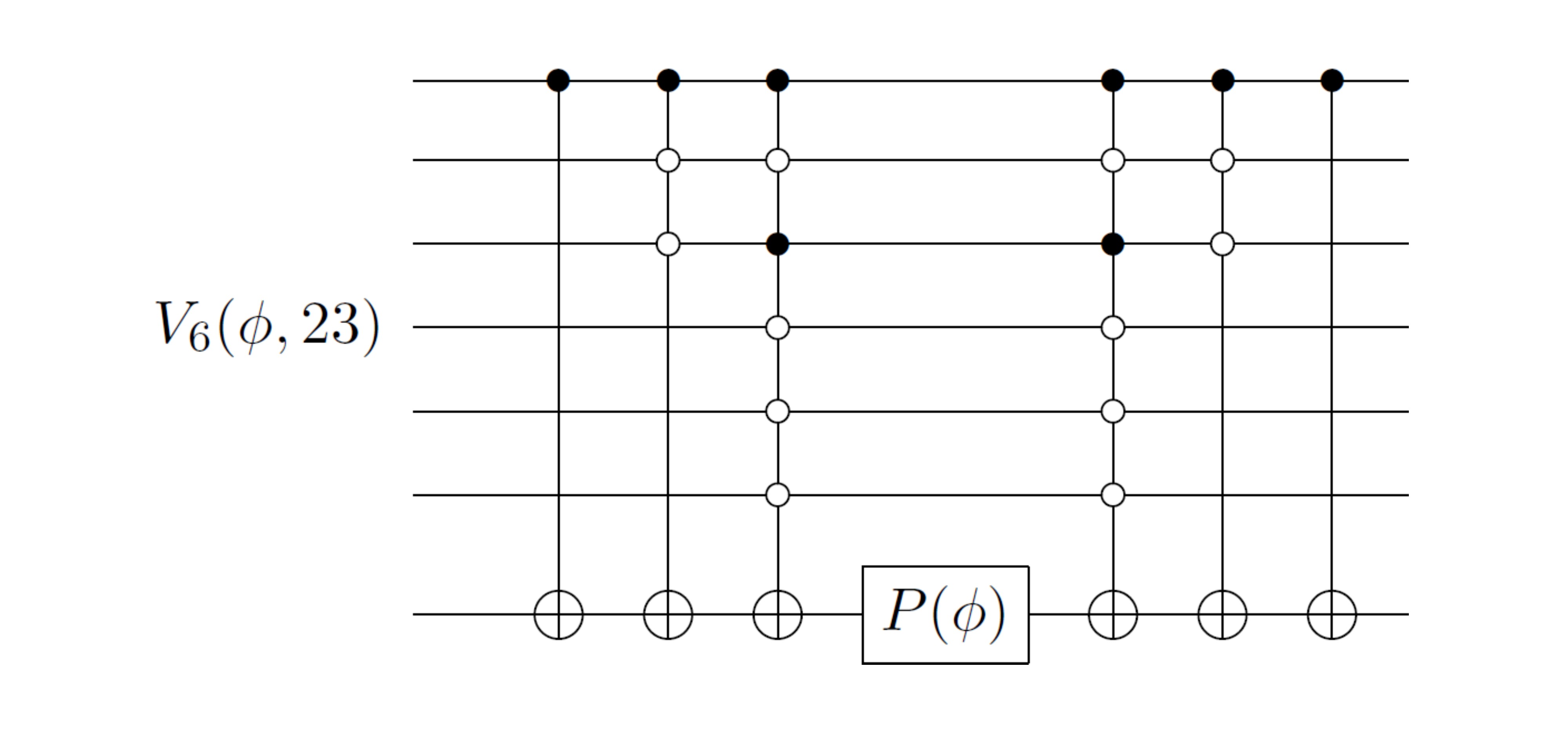}
\fcaption{\label{fig:V:6:23} A circuit for $V_6(\phi,23)$. }

}
\end{figure}

\begin{algorithm}[H]
\caption{\mbox{CED}: Decomposition of $X^n(\ell)$ into Toffoli gates}
\label{cpt:cascaded:entanglers}
\algsetup{indent=2em}
\begin{algorithmic}[1]
\REQUIRE{$n, \ell$}
\IF {$\ell=0$}
\RETURN {$\hat{I}_{n+1}$}
\ENDIF
\STATE {Set $2^{k_1}+\ldots+2^{k_h}$ binary decomposition of $\ell$}
\STATE {$j_r \gets 2^{n-k_r} - \sum_{j=r}^h 2^{k_j-k_r}, r = 1,\ldots,h$}
\STATE {$c_1 \gets \prod_{r=1}^h \,\Lambda^{n[j_r]}(X)$}
\STATE {Set $\ell_1 \, 2^{k_1}+\ldots+ \ell_h \, 2^{k_h}= \mbox{bsb}(\ell)$}
\STATE {$j_r \gets 2^{n-k_r} - \sum_{s=r}^h \ell_s \, 2^{k_s-k_r}, r = 1,\ldots,h$}
\STATE {$c_2 \gets \prod_{s=1}^h \,\Lambda^{n[j_r]}(X)$}
\IF {$Tcount(c_1) < Tcount(c_2)$}
\RETURN {$c_1$}
\ELSE
\RETURN {$c_2$}
\ENDIF
\end{algorithmic}
\end{algorithm}
\vspace*{12pt}
\noindent
\begin{example}\label{x6}
Use $\mbox{bsb}(23) = 32 - 8 - 1$ to decompose $X^6(23)$ into

$\Lambda^{6[1]}(X) \, \Lambda^{6[100]}(X) \, \Lambda^{6[101000]}(X)$ yielding a decomposition with three factors: the first is a CNOT and the other two have a total of $9$ levels of control.
\end{example}
Note for completeness that the ordinary binary expansion $23 = 16+4+2+1$ would have spawned a decomposition into four factors of $\Lambda^{n[j]}$ type and a total of $17$ levels of control.
For $\Lambda^{6[101000]}(X)$ it is advisable to use an additional ancilla qubit, and thus to implement it as an $8$-qubit circuit.
As per  \cite{BarencoEtAl1995}, Lemma 7.4, the implementation requires $8 \times (8-5) = 24$ three-qubit Toffoli gates and so the required $T$-count is bounded by $24 \times 7 = 168$.
$\Lambda^{6[100]}(X)$ is realized with four three-qubit Toffoli gates and no ancillas.
The corresponding circuit for implementing $V_6(\varphi, 23)$ is similar to the one presented in Fig.~\ref{fig:V:6:23}.

\subsection{The $\mbox{CED2}(x,r,x)$ Procedure}

The $\mbox{CED2}(x,r,x)$ procedure referenced in Algorithm \ref{alg:cascaded:entanglers}
can be implemented by concatenating a circuit $c$ for the entangler $x$ in its first argument obtained via Algorithm \ref{cpt:cascaded:entanglers} with a Clifford+T circuit for the rotation $r$ and then with the mirror image of the circuit $c$, as we have just done for $V_6(\varphi,23)$ in the above example.
However, in most cases this is not optimal and the synthesis of $V(\phi, \ell)$ benefits from further optimization of the entangling component.

\begin{figure}[htbp]
{
\centering
\includegraphics[scale=0.2]{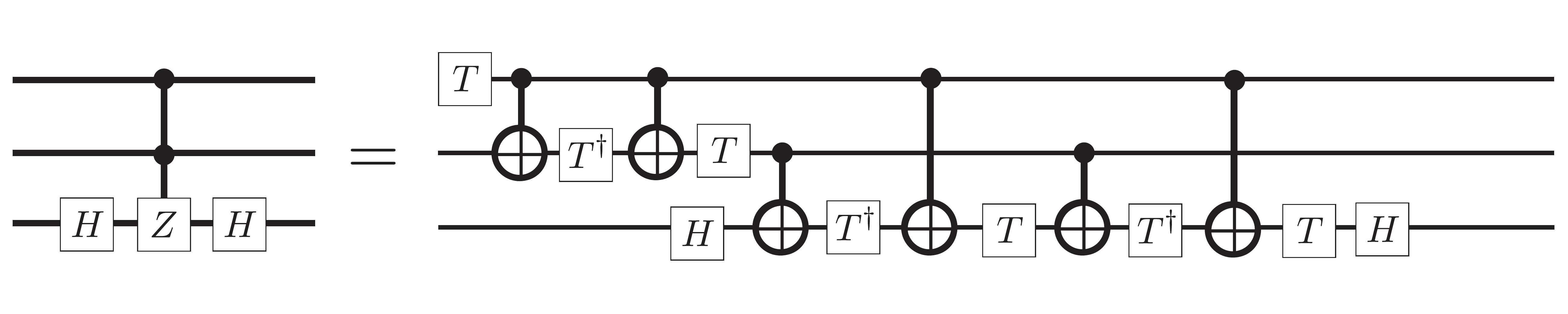}
\fcaption{\label{fig:cchzh:decomp} An exact decomposition of the Toffoli gate on the Clifford+$T$ basis. We use the fact that $HZH=X$, where $H$ is the Hadamard gate and $Z$, $X$ are the standard Pauli gates.}
}
\end{figure}

A pair of matching multi-controlled-NOTs (such as those in Figure \ref{fig:simple:cascading:entangler}) can be implemented at less than twice the cost of a single multi-controlled-NOT with the same number of controls.
Instead of implementing each of the multi-controlled-NOTs faithfully, we implement them up to mutually-cancelling local phases of $-1$ at a lower cost.
For example, Figure \ref{fig:cchzh:decomp} shows the exact Clifford+$T$ decomposition of the Toffoli gate using 7 $T$ gates.
Note that we have used the identity $HZH=X$ to rewrite the Toffoli as a double-controlled-$Z$ gate, which is diagonal and hence has an exact decomposition in the Clifford+$T$ basis using the technique of \cite{WelchJ2014}.\footnote{This is an interesting special instance of a circuit that under the Walsh decomposition technique produces a circuit directly into the Clifford+T basis in terms of precisely the minimal number of T gates required to implement it.}

\begin{figure}[htbp]
{
\centering
\includegraphics[scale=0.12]{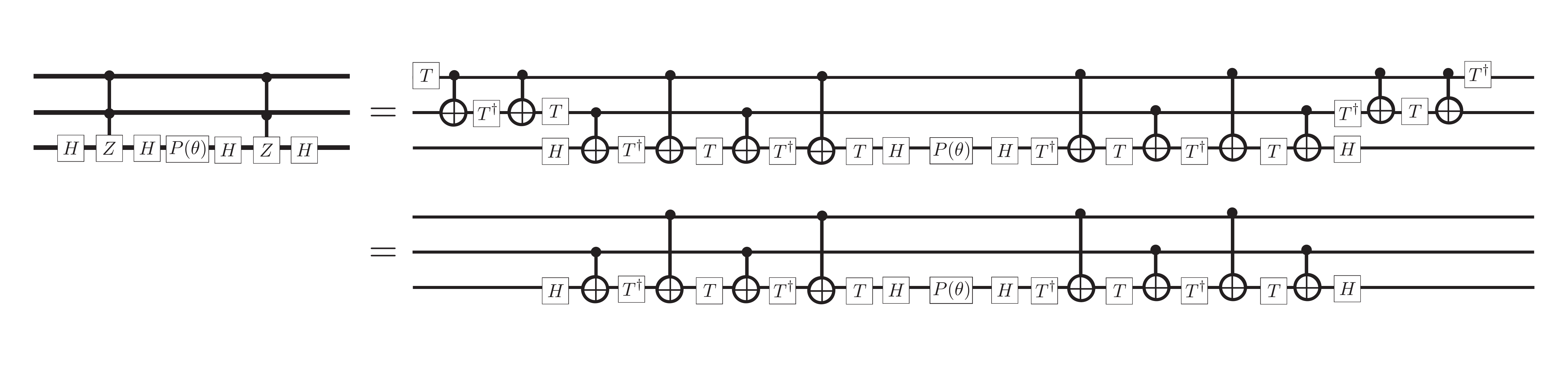}
\fcaption{\label{fig:pairedentanglers}An exact decomposition of two paired Toffoli gates separated by a phase gate on the target qubit. The 6 T gates and 4 CNOTs appearing on the upper 2 qubits can all be commuted to one side and hence cancel with one another. Reducing the total T-count of the pair of entanglers from 14 to 8.}
}
\end{figure}

Figure \ref{fig:pairedentanglers} shows an exact decomposition of two paired Toffoli gates with a phase gate on the target qubit. We use the Clifford+$T$ decomposition of the Toffoli shown in Figure \ref{fig:cchzh:decomp}.
The first two qubits in this example have a sequence of gates which are adjoints of one another.
Thus if we carry through all commutations to move the gates on the top two qubits towards the center, we find that not only do they cancel out, but all commutations in between different CNOT operators also mutually cancel.
Independently they would cost 7 $T$ gates each for a total of 14 per pair, however in this paired configuration a pair of matched Toffoli entanglers costs only 8 $T$ gates.

\begin{figure}[htbp]
{
\centering
\includegraphics[scale=0.1]{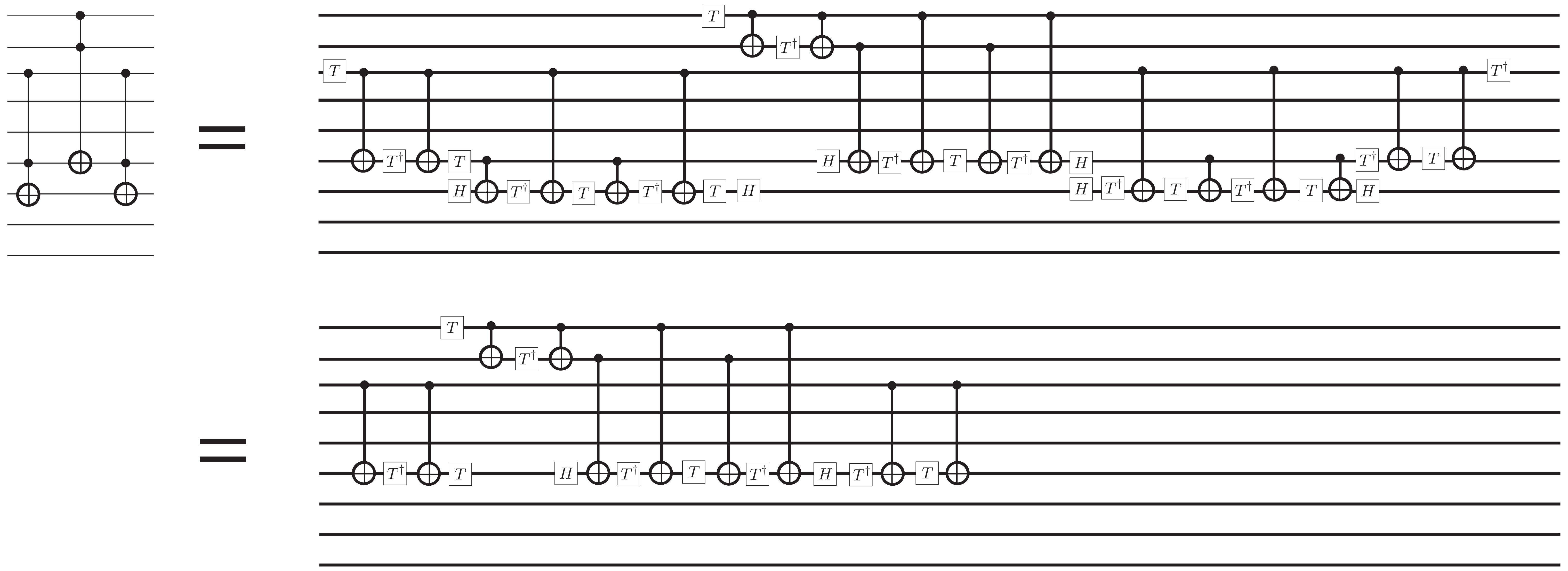}
\vspace*{10pt}
\fcaption{\label{fig:pairedentanglers2} A pair of Toffoli gates with a third Toffoli in between decomposed over the Clifford+$T$ basis using the same Toffoli decomposition technique as Figure \ref{fig:pairedentanglers}. There is no rotation $P(\theta)$ between the targets of the paired Toffolis allowing for an even greater simplification in $T$-count. This type of pattern appears often in the multi-controlled entangler decomposition proposed in \cite{BarencoEtAl1995} which is used as part of the CED algorithm. The pair of matched Toffoli gates requires only 4 $T$ gates (the other 5 in the circuit are for the center Toffoli).}
}
\end{figure}

Another type of simplification can occur as part of the multi-controlled entangler decompositions proposed in ~\cite{BarencoEtAl1995}.
 A typical pattern that appears in these decompositions is shown in Figure \ref{fig:pairedentanglers2}. In this case, there is no phase gate separating the target qubits of a pair of matched entanglers, and as a result a significant simplification occurs when we pair two Toffolis and examine them under the Clifford+$T$ decomposition of Figure \ref{fig:cchzh:decomp}.
As it turns out the two paired entanglers in this configuration require only 2 $T$ gates each, resulting in a total of 4 $T$ gates for the pair.
Several other types of simplifications may occur in these types of paired entangler scenarios. Overall, finding the true optimum for entanglement cost is beyond the scope of this paper and merits future work.

\section{Empirical Evaluation}
\label{sec:expts}

In this section we explore practical tradeoffs between using decomposition with cascaded entanglers and a Walsh-based decomposition for the approximate implementation of diagonal unitaries $U$.
Both types of decomposition perform a reduction of the target diagonal unitary $U$ to a collection of single-qubit axial rotations in the first stage. We assume that CNOT gates incur trivial costs in terms of their $T$-count in the resulting circuits. An exact Walsh-based decomposition uses only CNOT gates for entanglement and hence incurs a trivial \emph{entanglement cost}. Algorithm \ref{alg:cascaded:entanglers} generates non-Clifford cascaded entanglers and thus yields a circuit with a potentially significant entanglement cost $E[U]$ which is independent from any approximation precision $\varepsilon$ and depends only on the shape of the diagonal of $U$ (in particular the number of qubits $U$ is defined over). Since $E[U]$ is fixed for a given $U$, it becomes asymptotically trivial with $\varepsilon \rightarrow 0$ and thus the entanglement cost of our algorithm is, in general, favorable for a range of values of $\varepsilon$.
We explore the range for which our algorithm outperforms a Walsh-based decomposition in terms of $T$-count.

\begin{figure}[htbp]
\centering
\includegraphics[width=4.5in]{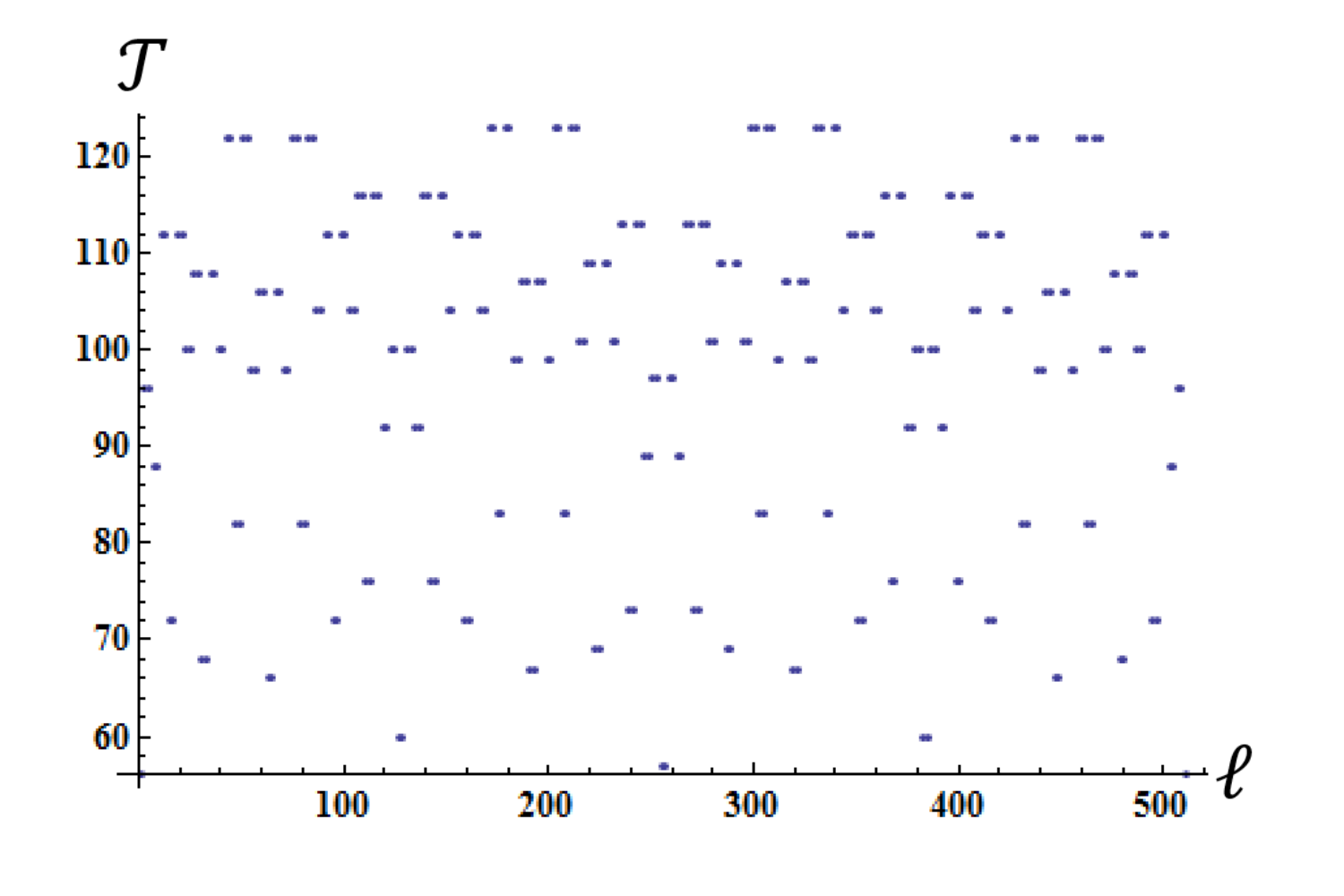}
\fcaption{\label{fig:tfc:plot}
Toffoli-count $\mathcal{T}$ of a $\mbox{CED}$ circuit on $n=10$ qubits for $X^n(\ell)$ versus $\ell$. (Balanced) Hamming weight of $\ell$ is the primary, but not the sole factor defining $\mathcal{T}$.}
\end{figure}

In order to define when our algorithm performs well, we need practical bounds on the entanglement cost $E[U]$ incurred by our algorithm. A numerical simulation demonstrates that the $T$-count of a Clifford+$T$ circuit generated by the $\mbox{CED}$ algorithm for an exact representation of the $X^n(\ell)$ operator exhibits a self-similar behavior as a function of $\ell$.
This behavior is similar to the behavior of the Hamming weight as a function of its argument.
A sample scatter plot of $\ell$ versus the Toffoli-count $\mathcal{T}$ for some 10-qubit unitaries is shown in Figure \ref{fig:tfc:plot}.
The distribution of Toffoli counts along the vertical axis is heavily skewed towards the worst case, (i.e., towards the maximum Toffoli-count).

We present our bound for the entanglement cost $E[U]$ based on the worst-case counts.
(Given the distribution it is easy to construct target diagonal unitaries with reasonably sparse phase contexts, such that all the cascaded entanglers $X^n(\ell)$ generated by our algorithm have near-worst-case cost.)
Based on numerical simulation, a good approximation for the worst-case count is of the form $\beta n^2$ where $\beta \sim 1.13$.

By design we have two identical cascaded entanglers of the form $X^n(\ell)$ per each of the phase factors (but the first one) that are to be matched in the target diagonal unitary.
Consequently, in the worst-case configuration each target phase factor incurs a $T$-count cost of
 $C_0  \log_2(1/\varepsilon) +  \kappa \, n^2$, where $C_0$ is the leading term maintained by a chosen method of approximating axial rotations over the Clifford+$T$ basis.
The Walsh-based decomposition generates a network with $2^n$ distinct rotations, except in special cases, and incurs a $T$-count cost of $2^n  C_0  \log_2(1/\varepsilon)$
\footnote{Note, again, that maintaining approximation cost in $O(\log(1/\varepsilon)$ is a relatively recent discovery (\cite{Selinger}, \cite{BoRoeSvoRUS}). With the less recent Solovay-Kitaev method (cf. \cite{DN}) the cost of single-qubit rotations would be more dramatic, and the choice would be heavily skewed in favor of the phase-context decomposition.}.

Therefore the condition for the cost boundary between the two methods is given by
\begin{equation} \label{eq:cost:parity:worst}
2^n  C_0  \log_2(1/\varepsilon) = (k - 1) (C_0  \log_2(1/\varepsilon) + \kappa\, n^2).
\end{equation}

\begin{figure}[htbp]
 \centering
\includegraphics[width=\textwidth]{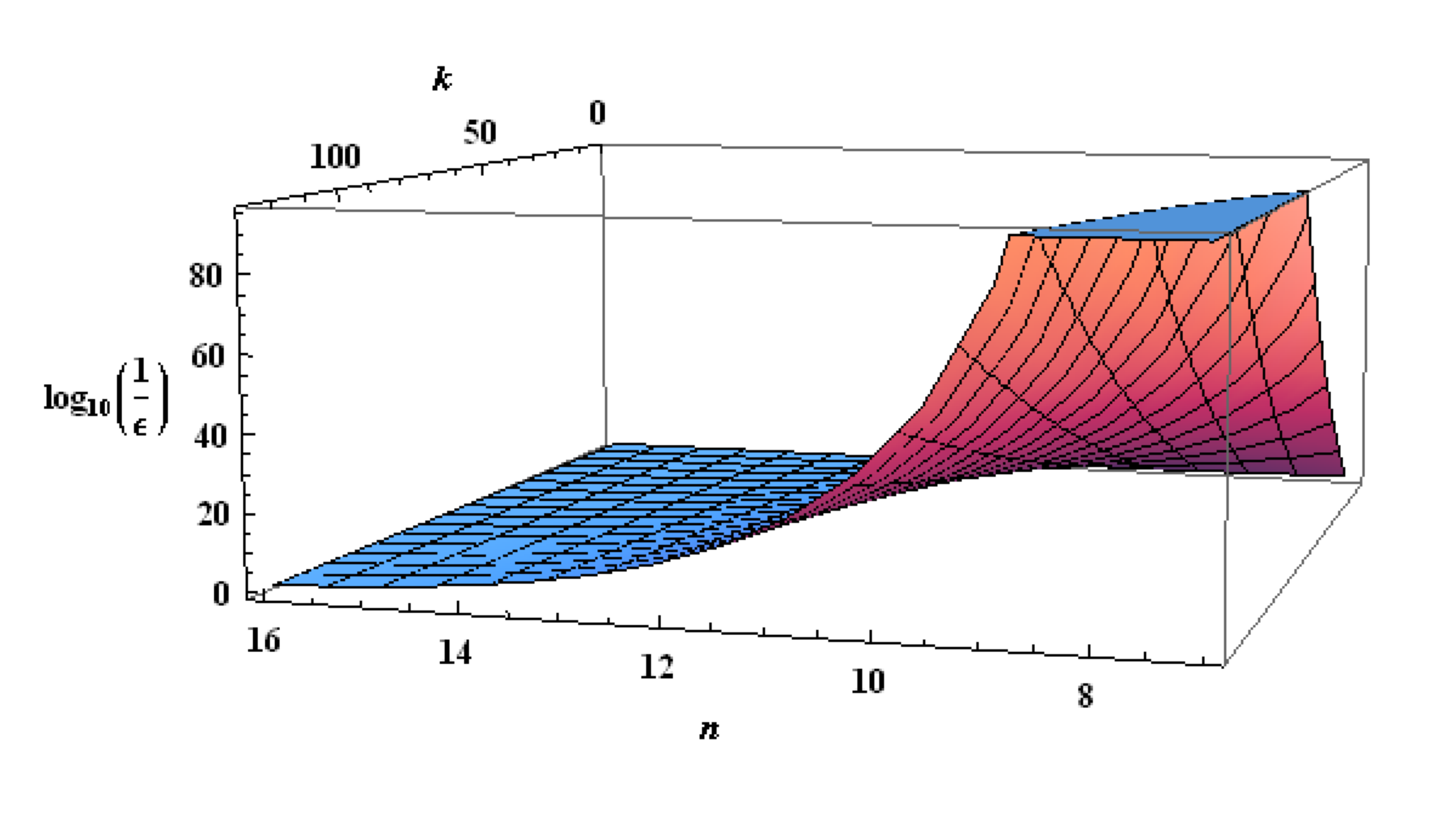}
\fcaption{\label{fig:decision:surface}
Decision surface in the worst case for cascaded-entangler versus Walsh-based decomposition for $k$ vs.~$n$ vs.~$\log_{10}(1/\epsilon)$. $(n,k,\varepsilon)$ triplets above the surface favor our CED algorithm; volume below the surface favors Walsh-based decomposition.}
\end{figure}

\begin{figure}[htbp]
\centering
\includegraphics[width=4.5in]{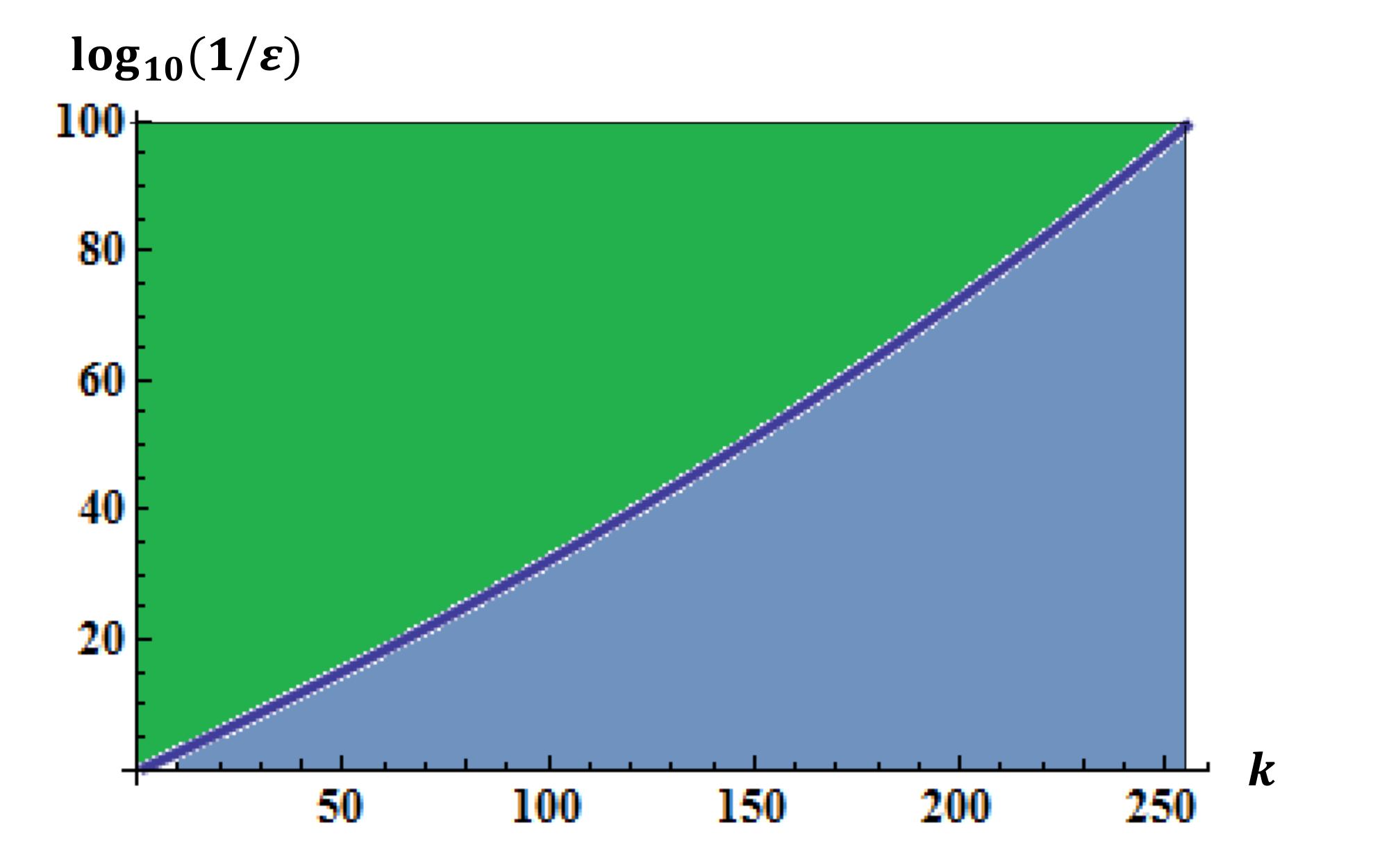}
\fcaption{\label{fig:cross:section}
Cross section of the decision surface for $n=10$ qubits. The upper (green)
area is populated with $(k,\varepsilon)$ pairs favoring our CED algorithm; the lower (blue) area favors
Walsh-based decomposition.}
\end{figure}

For the present evaluation, we estimate $C_0 \sim 1.15$ based on \cite{BoRoeSvoRUS} and $\kappa \sim 10$ based on \cite{GilSel}.
The corresponding decision surface in $n,k,\log_{10}(1/\varepsilon)$-space is shown in Figure \ref{fig:decision:surface}.  A cross-section for $n=10$ qubits is shown in Figure \ref{fig:cross:section}.
For $(n,k,\log_{10}(1/\varepsilon))$ triplets above the decision surface in Fig.~\ref{fig:decision:surface} and for a $(k,\log_{10}(1/\varepsilon))$ pair above the decision line in Fig.~\ref{fig:cross:section}, cascaded-entangler decomposition is favored over the Walsh-based decomposition.

Pre-selecting one of the two decomposition methods based on the worst-case bounds could be too pessimistic.
Indeed, Figure \ref{fig:tfc:plot} suggests that there is a significant share of configurations where the cost of cascaded entanglers required by our algorithm is relatively low.
Therefore we recommend performing two decompositions in parallel and selecting the circuit with lower cost.

\begin{figure}[htbp]
\centering
\includegraphics[width=4.5in]{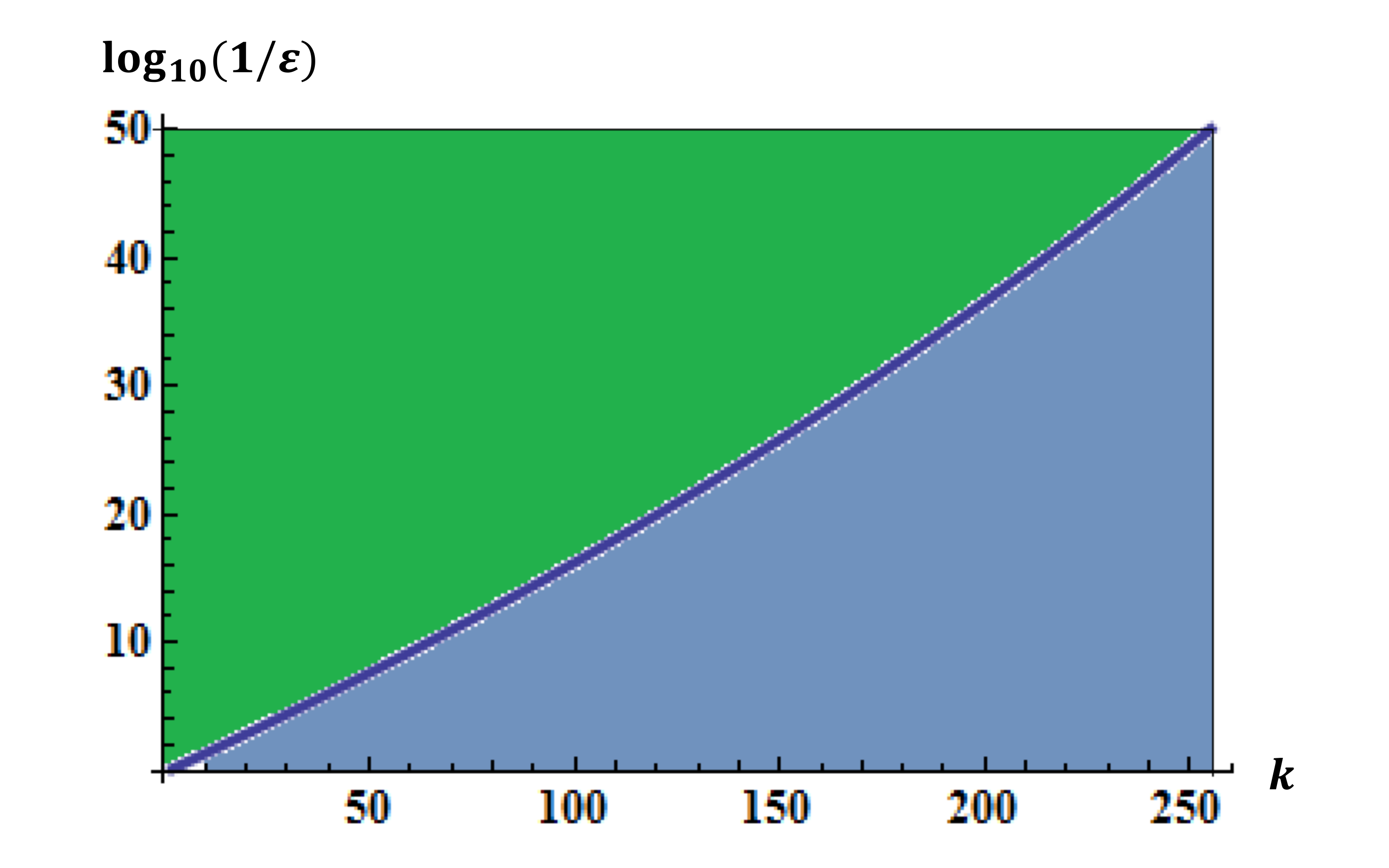}
\fcaption{\label{fig:favorable:cases}
Decision line for $n=10$ qubits, "best case" targets.The upper (green)
area is populated with $(k,\varepsilon)$ pairs favoring our CED algorithm; the lower (blue) area favors
Walsh-based decomposition.}
\end{figure}

We highlight the possible benefits of this approach by studying a set of favorable cases.
For each $n$ consider the set of odd values of $\ell$ smaller than $2^n$.
Let $m(n) = \min_{\ell} \{T\mbox{-count}(X^n(\ell))\, | \, \mbox{odd} \, \ell < 2^n\}$.
Numerical simulation and subsequent estimation suggests a linear fit
of $m(n) \sim 72\,(n-3)$ and an overall estimated cost of
$C_0  \log_2(1/\varepsilon) + 72\,(n-3)$ for each phase factor.
The boundary based on the cost of the two methods for this subset of target diagonal unitaries is
\begin{equation} \label{eq:cost:parity:best}
2^n  C_0  \log_2(1/\varepsilon) = (k - 1) \left(C_0  \log_2(1/\varepsilon) + 72\,(n-3)\right).
\end{equation}
The decision line assuming the best-case configuration of a $10$-qubit diagonal unitary target is shown in Figure \ref{fig:favorable:cases}.

To summarize, for a practically important range of target precision values
 $\varepsilon \in [10^{-5},10^{-40}]$ our algorithm for the decomposition of diagonal unitaries with cascaded entanglers produces significantly shorter Clifford+$T$ circuits for a wide range of sizes $k$ of the phase context even when the structure of the target diagonal is worst-case.
For a tiny phase context (single-digit $k$) the cost advantage of our circuits over the Walsh-based circuits is likely to be exponential in $n$.
 An analysis of good cases (corresponding to the troughs in the cascaded-entangler cost) suggests that our algorithm should be run in parallel with the Walsh-based one, and the more cost-effective circuit may then be chosen.

\section{Conclusions and Future Work}
\label{sec:concl}


We have analyzed the cost of approximating diagonal unitaries over the Clifford+$T$ basis.
We find that by distinguishing between a \emph{phase-sparse} and \emph{phase-dense} diagonal unitary, the $T$-count and cost of approximation can be significantly reduced.
Namely, if the target diagonal unitary is phase-sparse, then the cost is dominated by the cost of approximating single-qubit rotations. We have designed a circuit decomposition framework and have shown that, given a diagonal unitary where phase factors on the diagonal belong to a collection with $k$ distinct values, the asymptotic implementation cost over the Clifford+$T$ basis is dominated by the $(k-1) C_0  \log_2(1/\varepsilon)$ term when the desired approximation precision  $\varepsilon$ tends to zero (and where $C_0$ is the known cost factor in the approximation cost of a general single qubit rotation). Intuitively our results suggest that when a target unitary can be described by a relatively small number of parameters, then that unitary can be approximated at a cost asymptotically proportional to the number of parameters and $\log_2(1/\varepsilon)$.

In the extreme case of an $n$-qubit diagonal with $k=2$ distinct phase factors in general position,
our method generates a circuit with only \emph{one} rotation and certain cascaded entanglers. Setting aside the entanglement cost that does not depend on precision and considering $\varepsilon \rightarrow 0$ we observe asymptotical cost improvement by a factor of $\Theta(2^n)$ compared to decomposition methods that are oblivious to the phase context.

While we have focused on asymptotic optimality of the decomposition of circuits, an important open topic for further research is establishing exact bounds for the complexity of cascaded entanglers.
In the broader context of quantum circuit decompositions, the distinction between phase rotation and entanglement should be applied to the synthesis of circuits for other classes of content-sparse gates, beyond the class of diagonal unitaries. Block-diagonal unitaries would be a likely candidate for the next stage of this research.

\nonumsection{Acknowledgements}
\noindent
The Authors wish to thank Vadym Kliuchnikov, Martin Roetteler, Dave Wecker and Nathan Wiebe for useful discussions.

\end{document}